\documentclass[aps,twocolumn,amsmath,amssymb,superscriptaddress]{revtex4}

\usepackage{graphicx}
\usepackage{dcolumn}
\usepackage{bm}

\begin{document}

\title{Microscopic and Macroscopic Signatures of Antiferromagnetic Domain Walls }

\author{R. Jaramillo}
\affiliation{The James Franck Institute and Department of Physics, The University of Chicago,
Chicago, IL 60637}
\author{T. F. Rosenbaum}
\email{t-rosenbaum@uchicago.edu}
\affiliation{The James Franck Institute and Department of Physics, The University of Chicago,
Chicago, IL 60637}
\author{E. D. Isaacs}
\affiliation{Center for Nanoscale Materials, Argonne National Laboratory, Argonne, IL 60439}
\author{O. G. Shpyrko}
\affiliation{Center for Nanoscale Materials, Argonne National Laboratory, Argonne, IL 60439}
\author{P. G. Evans}
\affiliation{Department of Materials Science and Engineering, University of Wisconsin,
Madison, WI 53706}
\author{G. Aeppli}
\affiliation{London Centre for Nanotechnology and Department of Physics and Astronomy, UCL,
London, WC1E 6BT}
\author{Z. Cai}
\affiliation{Advanced Photon Source, Argonne National Laboratory, Argonne, IL 60439}

\date{\today}

\begin{abstract}
Magnetotransport measurements on small single crystals of Cr, the elemental antiferromagnet, reveal the hysteretic thermodynamics of the domain structure. The temperature dependence of the transport coefficients is directly correlated with the real-space evolution of the domain configuration as recorded by x-ray microprobe imaging, revealing the effect of antiferromagnetic domain walls on electron transport. A single antiferromagnetic domain wall interface resistance is deduced to be of order $5\times10^{-5}\mathrm{\mu\Omega\cdot cm^{2}}$ at a temperature of 100 K.   
\end{abstract}

\maketitle

Magnetic domains constitute the internal architecture of a host of technologically interesting materials. How ferromagnetic domains form, move and scatter electrons lies at the heart of items from electrical motors and transformers to data storage devices \cite{Krusin-Elbaum2001}. In an ordinary ferromagnet (FM), a domain is characterized by a single vector, namely its magnetization. Antiferromagnets (AFM) typically are characterized by multiple vectors corresponding to the local magnetization and how it evolves with position, and offer new and expanded microscopic architectures for exploitation. However, with neither a net magnetic moment nor long wavelength features, antiferromagnetic domains have resisted the detailed characterization that underpins the applications prevalent for ferromagnetic domains. As the ability to craft device features progresses to ever smaller dimensions, and hybrid devices mixing ferromagnetic and antiferromagnetic components proliferate \cite{Nolting2000}, the need to understand and manipulate antiferromagnetic domains on the microscale, and domain walls on the nanoscale, becomes increasingly acute.

We present here a combined electrical transport and x-ray microprobe imaging study of a model AFM. We show that magnetotransport measurements on small single crystals of Cr, the elemental spin-density-wave AFM, are highly sensitive to the domain structure, and we directly correlate the temperature-dependent and hysteretic behavior of the transport coefficients with the real-space evolution of the domain structure. Combining the x-ray images with the measured anisotropic resistivity yields a quantitative estimate of the electrical resistance of a single antiferromagnetic domain wall. 
 
\begin{figure}
\includegraphics{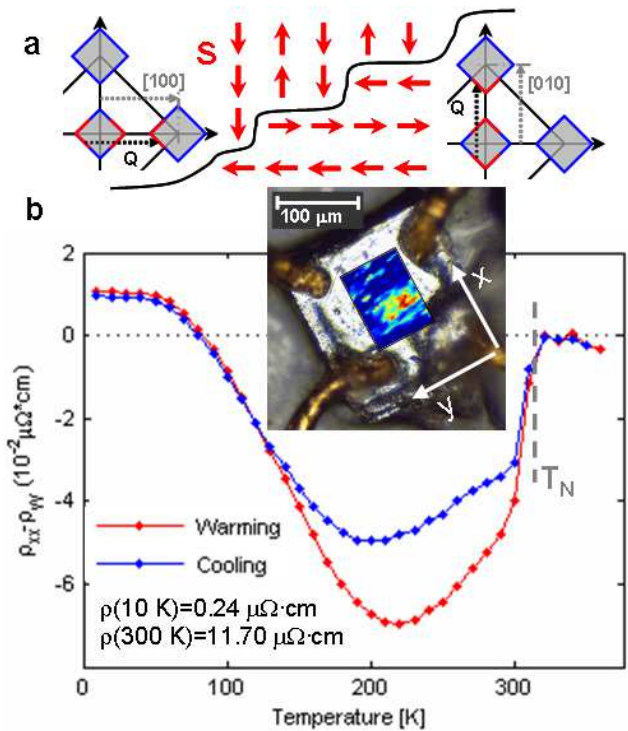}
\caption{
\label{fig-first} (a) Schematic AFM domain structure in Cr. Only one of the two possible spin polarizations $\mathbf{S}$ is illustrated for each Q-domain; hence we show only one of the four types of boundary that are possible between this pair of Q-vectors. In reciprocal space, the gapped and ungapped Fermi surfaces are shown in red and blue, respectively. The electron Fermi surface centered at the middle of the Briulloin zone (BZ) and the hole surface centered at the corner of the BZ are connected by the nesting vector $\mathbf{Q}$. Only the magnetic bands are shown for clarity. (b) Difference between the longitudinal resistivities $\rho_{xx}$ and $\rho_{yy}$. The coordinates are defined by the geometry of our electrical measurement and are described in the inset. (Inset) Superposition of actual crystal wired for electrical measurements with a typical Q-domain image to scale, permitting direct comparison of the length scales involved. Domain wall motion, such as that seen in Figs.~\ref{fig-second} and~\ref{fig-third}, is significant on the length scale of the current paths.
}
\end{figure}

The incommensurate spin-density-wave (SDW) state in chromium is a partially-gapped electron-hole pairing state that is caused by a nesting instability of the paramagnetic Fermi surface \cite{Fedders1966,Werner1967}. The SDW modulation vector $\mathbf{Q}$ is selected by the nesting condition and may lie along any of the cubic crystallographic axes, leading to so-called ``Q-domains.'' Between the N\'{e}el ordering temperature $T_N=311\:\mathrm{K}$ and the spin-flop temperature $T_{SF}=123\:\mathrm{K}$, the SDW is in the transverse phase ($\mathbf{S}\perp\mathbf{Q}$) and the spins preferentially lie along either cubic axis perpendicular to Q, leading to so-called ``S-domains.'' Below $T_{SF}$ the SDW is longitudinal ($\mathbf{S}\parallel\mathbf{Q}$). This multiple degeneracy, threefold for the direction of $\mathbf{Q}$ and twofold for the direction of $\mathbf{S}$ in the transverse phase, leads to a rich variety of domain interfaces involving rotations of both the Fermi surface and the spin polarization, with great potential for modulating spin and charge transport.

The partially-gapped nature of the SDW state has complicated anisotropic effects on transport, which can be modeled semi-quantitatively. Results for the resistivity tensor are available in the literature \cite{Trego1968,Furuya1976}, derived from samples that have been specially prepared in a single-Q-domain state, with a resistivity anisotropy of approximately 10\% at low temperature. However, akin to the state of play for ferromagnets in the 1960s \cite{Taylor1968,Cabrera1974}, little is known for antiferromagnets about the effect of domain walls on the electrical transport. 

We present in Fig.~\ref{fig-first}(a) a schematic AFM domain structure. The domain wall (DW) is defined by two potentially independent rotations $\hat{\eta}_{S}/R$ and $\hat{\eta}_{Q}/R$ of the spin polarization and modulation vectors, respectively, where $R$ represents the wall thickness. This compares to a FM where the DW simply is defined by a single rotation $\hat{\eta}_{S}/R$. The electronic properties of the DW are dominated by the ability of the electrons to scatter between domains with differing Fermi surfaces, and this is largely dependent on the relative extents of the domain wall $R$ and the conventional electron mean free path $l$ . For domain walls in a conventional FM such as Co, $R/l\gg1$, with the result that DW interface resistances are small ($\sim10^{-7}\:\mathrm{\mu\Omega\cdot cm^{2}}$) \cite{Lee2006}. For $R/l\gg1$, quantum effects become important and in materials with sharp DW features the resulting tunneling magnetoresistances can be much larger ($\sim10^{-6}-10^{-5}\:\mathrm{\mu\Omega\cdot cm^{2}}$) \cite{Mathur1999}. At Q-domain walls in Cr there is an abrupt, several lattice-constants wide\cite{Fenton1980,Braun2000} range over which the anisotropic gap in the Fermi surface rotates by $90^{o}$ from one of the cubic axes to another.  This is accompanied by the observed $90^{o}$ rotation in the spin-density, charge-density, and lattice strain modulation \cite{Braun2000,Evans2002}. This means that when electrons flow across a DW they move from an ungapped metallic Fermi surface to one that is gapped and insulating. The condition $R/l\sim1$ is satisfied in Cr, where both quantities are on the order of a few nm \cite{Braun2000}. Q-domain walls, therefore, may be significant charge- (and spin-) dependent scatterers. 

We prepared three crystals measuring $(195\times180\times45)\:\mathrm{\mu m^{3}}$, $(460\times475\times60)\:\mathrm{\mu m^{3}}$, and $(675\times695\times80)\:\mathrm{\mu m^{3}}$ for transport measurements, guided by the work of Evans et al. \cite{Evans2002} who showed that the length scale of the Q-domains is tens of microns on a side, with S-domains somewhat smaller. Our intent was to measure samples small enough so that the movement of a few domain walls would cause a measurable change in the resistivity (see Fig.~\ref{fig-first}(b), inset), but large enough to remain fully in the bulk regime. All samples were oriented Cr single-crystals cut along the non-cubic $(\bar{1},\bar{1},2)$, $(1,\bar{1},0)$, and $(1,1,1)$ planes, polished to an optical finish, and etched to reduce domain pinning by surface anisotropy \cite{Braun2000}, surface roughness and crystallographic strain. The full resistivity tensor in the plane of the measurement was measured in the van der Pauw configuration \cite{Montgomery1971}. All temperature changes were performed in zero-field to avoid field-induced biasing effects. Magnetic field measurements were in the linear regime with $H\leq 0.5\:\mathrm{T}$.

\begin{figure*}
\includegraphics{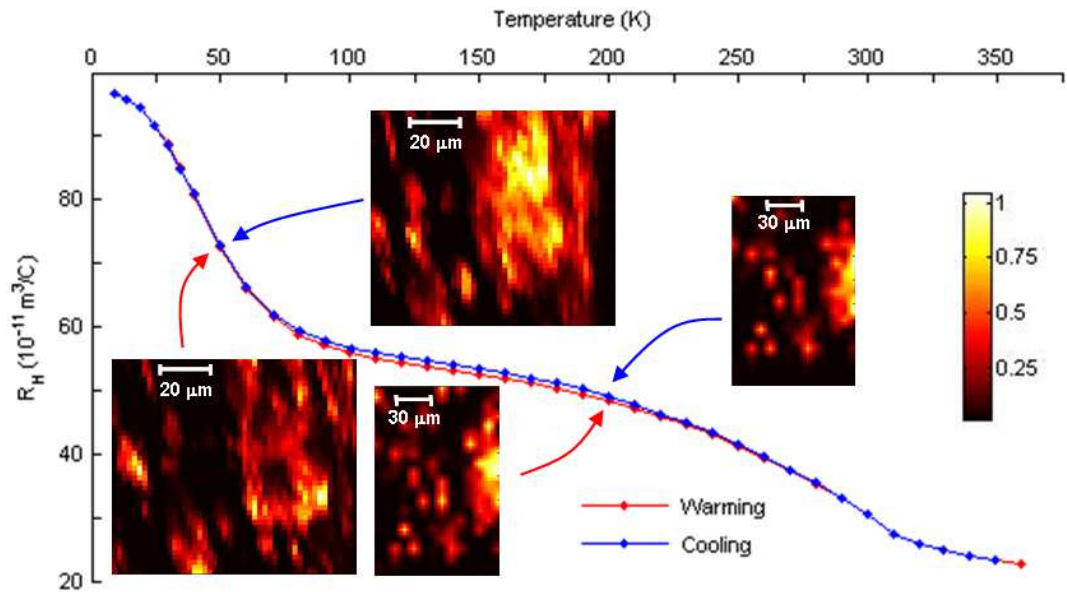}
\caption{
\label{fig-second} Thermal hysteresis of the Hall coefficient measured on a sub-mm Cr crystal. Images are maps of microprobe diffraction intensity from one of three Q-domain types. Beam spot size is $300\times600\:\mathrm{nm^{2}}$, diffraction is from a $(111)$ face, x-ray energy is $11.6\:\mathrm{keV}$, and the penetration length is $\approx1.5\:\mathrm{\mu m}$. Images are taken at $50\:\mathrm{K}$ and $200\:\mathrm{K}$ within a thermal cycle between $50\:\mathrm{K}$ and $300\:\mathrm{K}$. Colorbar indicates diffraction intensity in cts/sec; multiplication factors of $2\times10^{4}$ and $10^{4}$ should be used for the $50\:\mathrm{K}$ and $200\:\mathrm{K}$ images, respectively. Images at the same T show the same sample area, but for different T the areas imaged are different.
}
\end{figure*}

\begin{figure}[h]
\includegraphics{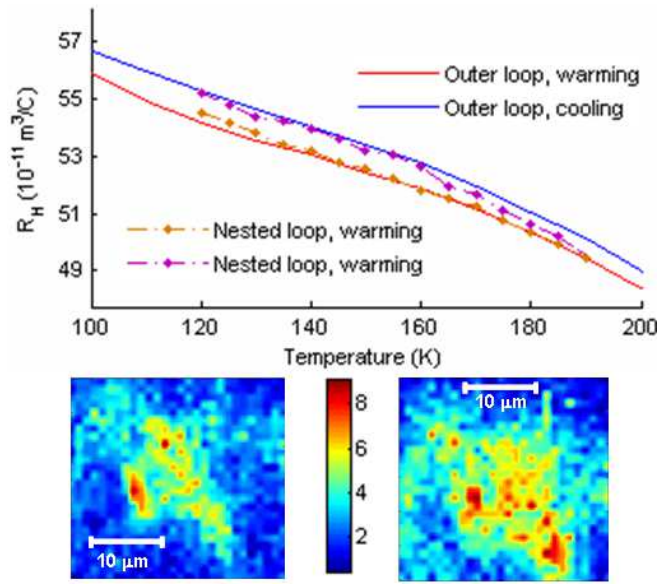}
\caption{
\label{fig-third} Hall coefficient around a nested temperature loop, showing persistence of hysteresis. Images are maps of microprobe diffraction intensity from one Q-domain type, taken at $110\:\mathrm{K}$ within a thermal cycle between $110\:K$ and $170\:K$, and show diffraction intensity from one of three Q-domain types. Beam spot size is $300\times600\:\mathrm{nm^{2}}$, diffraction is from a $(111)$ face, x-ray energy is $5.8\:\mathrm{keV}$, and the penetration length is $\approx2\:\mathrm{\mu m}$. Color bar indicates diffraction intensity in cts/sec. Images show the same sample area.
}
\end{figure}

The transport measurements for the smallest sample are summarized in Figs.~\ref{fig-first}-\ref{fig-third}. In Fig.~\ref{fig-first}(b) we plot the difference $\rho_{xx}-\rho_{yy}$ between the zero-field longitudinal resistivities in the sample plane, an effective rotation of the scattering, for both warming and cooling. Particularly striking is the way in which this difference suggests a shifting domain configuration with temperature, with a clear onset just below $T_{N}$. The differential data also exhibit a pronounced thermal hysteresis, which is large compared to the hysteresis present in the bare longitudinal resistivities.

We find thermal hysteresis in all of the measured resistivity components, but the effect is largest in the Hall coefficient (see Figs.~\ref{fig-second} and~\ref{fig-third}). The Hall effect's privileged position as a sensitive indicator of domain structure is mirrored by its sensitivity to the onset of the SDW itself at the quantum critical point \cite{Yeh2002}. The hysteresis loop is robust, repeating over many thermal cycles spanning hundreds of hours. The lower and upper temperatures that define the Hall hysteresis are $75\pm15\:\mathrm{K}$ and $250\pm 15\:\mathrm{K}$. There is no signature of the spin-flop transition in the transport data, pointing to the Q-domains rather than the S-domains as the source of the hysteretic behavior. Measurements using thermal equilibration times differing by more than an order of magnitude did not affect the observed response and we saw no evidence of glassy relaxation or aging over hours. 

We plot our ``master'' Hall curve in Fig.~\ref{fig-second}. The response follows the master curve regardless of whether the uppermost temperature is above (as shown) or below (not shown) $T_{N}$. Notably, the response remains on the master curve as the system undergoes a series of nested thermal loops. The innermost such loop is shown in Fig.~\ref{fig-third}. We find that the system doesn't immediately snap to the cooling master curve after the turnaround point but does indeed find this master curve after a few downward steps in temperature; it exhibits macroscopic return point memory \cite{Pierce2003}. These results suggest that the Hall coefficient is particularly sensitive to the underlying domain structure. Above some "fixing" temperature the preferred domain distribution depends on whether the temperature has been increasing or decreasing; below this fixing point the distribution may settle into a single configuration regardless of the direction of temperature change. We note that the upper bound of our hysteresis loop corresponds to the temperature at which Q-domain fluctuations are no longer detectable in electrical noise measurements \cite{Michel1991}.

In order to make connections to the underlying microscopic physics, we took crystals from the same wafer to the x-ray microprobe beamline 2ID-D at the Advanced Photon Source and imaged the AFM domains using a sub-micron focused beam \cite{Evans2002}. The resulting images are presented in Figs.~\ref{fig-second} and~\ref{fig-third}, mapped onto the hysteresis loop defined by the Hall coefficient. We show in Fig.~\ref{fig-second} pairs of Q-domain images taken at 50 K and at 200 K, near the edges of the measurable hysteresis as the system executed a round-trip temperature cycle between 50 and 300 K; diffraction is from the CDW satellite at $(0,0,2Q)$, and therefore is sensitive to only one of the three types of Q-domain (that with $\mathbf{Q}\parallel[001]$). The domain patterns on warming differ from those taken at the same temperature on cooling, and the nature of these differences provides insight into the physical mechanism underlying the hysteresis in the electrical measurements. This is seen most clearly in the two images taken at 50 K, a temperature at which there is no hysteresis in the transport data. Although the interior structures - the relative Q-domain populations - change, it is apparent that most of the domain walls have returned to their same positions. This effect may be quantified by comparing the changes in total domain area and boundary length between pairs of images taken at the same temperature. We define the extent of a domain by the condition that the diffraction intensity be at least half of the peak intensity measured at the domain centers. Comparing the 50 K images, the volume occupation of the observable Q-domain type for this scattering geometry has changed by 63\% on cooling as compared to warming, but the change in domain wall length is only 4\%. At 200 K, where the Hall effect still demonstrates hysteresis, the volume occupation differs by 48\% and the domain wall length has changed by 42\%. It is the spatial distribution of domain walls - not the fractional volume of occupation of the Q-domains - that appears to be most strongly selected by the pinning landscape and that correlates with the hysteresis. We conclude that the domain walls themselves have a measurable effect on transport, and indeed have the dominant effect on the hysteresis.

We focus in Fig.~\ref{fig-third} on images taken at 110 K, near the widest part of the hysteresis loop. Diffraction is from the magnetic satellite at $(0,1-Q,1)$; below $T_{SF}$ entire Q-domains with $\mathbf{Q}\parallel[010]$ diffract at this position \cite{Evans2002}. Comparing the images we see that there is significant hysteresis in the domain configuration, with a 182\% change in the volume occupation and an 81\% change in the domain wall length. This large change is consistent with the Hall hysteresis, which reaches its maximum close to 110 K. 

The measured effect of domain motion on transport should decrease as the crystal size is increased since the longer current paths will see a greater number of domain walls and the effects of individual domain wall motion should then average out. This hypothesis is borne out by our measurements on the series of crystals of increasing size. As the sample volume increases by a factor of eight, and then by another factor of three, the hysteresis in the Hall effect decreases by 35\%: from a maximum of 2\% to 1.7\% to 1.3\%. 

We can estimate the resistance of a single domain wall by comparing results for the resistivity anisotropy in single-Q samples with data taken on poly-Q samples. This works best for bulk crystals where many domains contribute to the scattering.  Taking values from the literature \cite{Trego1968}, we solve for an effective domain wall contribution to the bulk resistivity that increases from of order $50\:\mathrm{n\Omega\cdot cm}$ at 100 K to $130\:\mathrm{n\Omega\cdot cm}$ at 200 K. However, bulk crystals do not permit a reliable estimate of the contributions of a single domain wall.  In our crystals, where there are only a few domains and we are able to determine an average domain length scale ($10\:\mathrm{\mu m}$) from direct imaging, we can take the bulk results and deduce a single AFM domain wall interface resistance of order $5\cdot10^{-5}\:\mathrm{\mu\Omega\cdot cm^{2}}$ at 100 K. This compares to the $R/l\sim1$ limit in FM, an intuitive result given the abrupt transition from ungapped to gapped Fermi surfaces across an AFM wall in Cr. A first principles theory of carrier scattering from AFM domain walls, involving much bigger symmetry groups than the analogous theory introduced decades ago for FM Bloch walls \cite{Cabrera1974}, would be an especially useful development. Our work represents the beginning of the science of AFM domain walls as elements in electronic devices, and is complementary to ongoing work on FM domains and domain walls \cite{Krusin-Elbaum2001,Nolting2000,Tang2004}. The unique advantages of antiferromagnetic domains, most notably the absence of an external magnetic field which makes them intrinsically less susceptible to stray fields and each other, can now also be harnessed for spintronics.

\begin{acknowledgments}
We are indebted to Leslie Sanford for assistance with sample preparation. The work at the University of Chicago was supported by the National Science Foundation, Grant No. DMR-0534296. The Advanced Photon Source is supported by the U. S. Department of Energy, Office of Science, Office of Basic Energy Sciences, under Contract No. W-31-109-Eng-38. R.J. gratefully acknowledges the support of a NSF Graduate Research Fellowship.
\end{acknowledgments}

\end{document}